\documentclass[graybox]{svmult}

\usepackage{mathptmx}       
\usepackage{helvet}         
\usepackage{courier}        
\usepackage{type1cm}        
\usepackage{makeidx}         
\usepackage{graphicx}        
\usepackage{multicol}        
\usepackage[bottom]{footmisc}

\usepackage{amsmath }
\usepackage{amsfonts}
\usepackage{amssymb}

\newcommand{\be}{\begin{equation}}
\newcommand{\ee}{\end{equation}}
\newcommand{\sech}{\mathrm{sech}}
\newcommand{\cotanh}{\mathrm{cotanh}}
\newcommand{\PT}{\emph{PT}}

\makeindex             

\begin{document}


\title*{Exactly solvable Wadati potentials in  the \emph{PT}-symmetric Gross-Pitaevskii equation}

 \titlerunning{ Exactly solvable Wadati potentials}
\author{I. V. Barashenkov, D. A. Zezyulin,   and V. {V}. Konotop}
\institute{I. V. Barashenkov 
\at National Institute for Theoretical Physics, Western Cape, South Africa and  Department of Mathematics, University of Cape Town,  Rondebosch 7701. \email{Igor.Barashenkov@uct.ac.za}
\and  D.  A. Zezyulin \at Centro de F\'isica Te\'orica e Computacional and Departamento de F\'isica,
Faculdade de Ci\^encias da Universidade de Lisboa, Campo Grande, Edif\'icio C8, Lisboa  P-1749-016, Portugal. \email{dzezyulin@fc.ul.pt}      
 \and V. V. Konotop \at Centro de F\'isica Te\'orica e Computacional and Departamento de F\'isica,
 Faculdade de Ci\^encias da Universidade de Lisboa, Campo Grande, Edif\'icio C8, Lisboa  P-1749-016, Portugal.       \email{vvkonotop@fc.ul.pt}}     
%
%
\maketitle

\abstract{
This note examines  Gross-Pitaevskii equations with \PT-symmetric potentials of the Wadati type: $V=-W^2+iW_x$.
We formulate a recipe for the construction of 
Wadati potentials supporting exact localised solutions.
The general procedure is exemplified by  equations with attractive and repulsive cubic nonlinearity bearing 
 a variety of bright and dark solitons.
}

\section{Introduction}
\label{Intro}

 A mean-field description of bosons with pairwise interaction is furnished by the Gross-Pitaevskii equation.
 In the one-dimensional geometry, the equation reads
 \be
i  \, \Psi_t + \Psi_{xx} - V(x)
 \Psi + g\Psi|\Psi|^2 = 0.
\label{0}
\ee

In this contribution, we will be concerned with the Gross-Pitaevskii  equations featuring 
complex potentials 
 $V(x)$~\cite{Muga,Moiseev}.
 In quantum physics, 
complex potentials provide a simple means to account for the inelastic scattering of particles
as well as for the 
 loading of particles in an open system~\cite{Cartarius,Cartarius1}.  The
 $x$-intervals  with $\mathrm{Im} \,V(x)>0$ and $\mathrm{Im} \, V(x)<0$ correspond to the gain and loss of particles, respectively.
When the gain exactly compensates the loss, that is, when $V$  obeys the symmetry
\be
\label{V_PT}
V^*(x)= V(-x),
\ee
the potential is referred to as the parity-time (\PT-) symmetric~\cite{Bender}. 

The nonlinear Schr\"odinger equation~(\ref{0}) with a $\PT$-symmetric potential (\ref{V_PT})  
is also used in the paraxial nonlinear optics~\cite{Ziad}.  In the optics contest, $t$ and $x$ 
stand for  the longitudinal and transverse coordinates, and $V(x)$ models the complex refractive index~\cite{Muga2}. 
	
Stimulated by the interest from the atomic physics and optics, 	
a number of exactly-solvable Gross-Pitaevskii equations was identified, both within and outside the \PT -symmetric variety. 
The list includes   periodic complex potentials~\cite{Musslimani2,AKSY}; the $\PT$-symmetric Scarff II~\cite{Ziad,Shi} and Rosen-Morse II potentials~\cite{Midya}, as well as a $\PT$-symmetric double-well superposition of a quadratic  and a gaussian~\cite{Midya2}.  

   
This contribution deals with potentials of the form    
\be
V(x)=-W^2+iW_x,
\label{Wadati}
\ee
where $W(x)$ is  a real function (called the {\it potential base\/} below), with $W(x) \to \mathrm{const}$ as $|x| \to \infty$.
Wadati was apparently the first who noted the relevance of potentials (\ref{Wadati}) for  the $\PT$-symmetric quantum mechanics~\cite{Wadati}.\footnote{
 Yet  these  have not been unheard of before.
 For instance, the  potentials \eqref{Wadati} appear in the context of supersymmetry
 \cite{Unanyan92,Balantekin91,Balantekin07}  and have applications in subatomic physics
 (where they were utilized for the modeling of 
neutrino oscillations  \cite{Balantekin98}).}
For the purposes of this study, 
we will be referring to \eqref{Wadati} as  the  Wadati potentials.

We consider the standing-wave solutions  $\Psi(x,t)=\psi(x)e^{i \kappa^2 t}$, where
$\kappa^2$ is real while the spatial part of the eigenfunction
  obeys the stationary equation
  \be
-\psi_{xx} +  V(x)
 \psi -g  \psi|\psi|^2 = -\kappa^2 \psi.
\label{A0}
\ee
In the linear case ($g=0$),
the  stationary Schr\"odinger equation with the potential \eqref{Wadati} and eigenvalue $-\kappa^2$
  can be mapped   onto the  Zakharov-Shabat  spectral problem, with the potential  $W(x)$
  and eigenvalue $i\kappa$
   \cite{Wadati74, Andrianov99,Lamb}.
This correspondence allows one to obtain  complex Schr\"odinger potentials with an entirely real spectrum
from the real Zakharov-Shabat potentials  whose entire discrete spectrum is  pure imaginary.
Potentials of the latter type are abundant --- in fact, all  Zakharov-Shabat 
eigenvalues of 
any 
 single-peaked real potential  $W(x)$  are pure  imaginary \cite{Klaus02, FatkOL14}.
 An example  of a multihump potential with an entirely imaginary discrete 
 spectrum is given by  the modified Korteweg-de Vries multisoliton 
 \cite{Wadati82}. 
 
In the nonlinear domain, the Gross-Pitaevskii equations with Wadati potentials enjoy an equally exceptional status. In the context of systems with gain and loss,
  the \PT-symmetric Wadati potentials are unique among all \PT-symmetric potentials 
  in supporting continuous families of  {\it asymmetric\/}  solitons \cite{Yang14d}.   
  This feature has an analogue outside the realm of \PT-symmetric systems.
 Namely, unlike the generic non-\PT~symmetric potentials, 
 the \PT-asymmetric Wadati potentials  bear   continuous families of stable nonlinear modes \cite{FatkOL14, KZ14b}.
(Generic non-\PT~symmetric complex potentials can only support  isolated dissipative solitons  rather than  continuous families of those \cite{AA05}.)
These unique attributes of the Wadati potentials 
stem from the fact that the stationary nonlinear Schr\"odinger equation (\ref{A0}) with $V$ as in
\eqref{Wadati}
has an $x$-independent invariant
\cite{KZ14b}.

  Finally, it is fitting to note  that the  Wadati potentials support  constant-density waves. 
  This property has been  used to study the modulational instability  within the Gross-Pitaevskii  equations with complex potentials \cite{MMCR15}.

In this contribution we propose a new procedure for the systematic
construction of exactly solvable Wadati potentials.  Here, we restrict ourselves to the \PT-symmetric case, that is, to the even functions $W(x)$.

Our approach is formulated in sections \ref{General}  and \ref{Dark} for the 
attractive ($g>0$) 
and repulsive ($g<0$) boson gas, respectively. 
The general procedure for the attractive nonlinearity is exemplified  by  two Wadati potentials with 
exact bright solitons (section \ref{2ex}).
In the repulsive-gas situation, we construct  potentials bearing exact 
 lump and bubble solutions (section \ref{kinks}). Finally, section \ref{flow} presents a Wadati potential generating a stationary flow 
of the condensate.

\section{General procedure: attractive nonlinearity}
\label{General}

We start with  the attractive nonlinearity,  $g>0$,
and assume that the potential has been gauged so that 
$W_x \to 0$ as $|x| \to \infty$.
Our main interest is in localised solutions; these obey 
\be
|\psi(x)| \to 0,  \
|\psi_x(x)| \to 0
\quad \text{as} \ |x| \to \infty.
\label{BC00}
\ee
The boundary conditions \eqref{BC00} require that $\kappa^2>0$. 
We let $\kappa>0$, for  definiteness.

It is convenient to  cast 
  the equation \eqref{A0} 
  in the form 
\be
u_{zz}   + (A^2  - i A_z)
 u + 2u|u|^2 =  u,
\label{A1}
\ee
where 
\[
A(z)= \frac{W(x)}{\kappa},
\quad
u(z)= \sqrt{\frac{g}{2}} \frac{\psi(x)}{\kappa},
\quad
z=\kappa x.
\]
The boundary conditions \eqref{BC00} translate into
\be
|u(z)| \to 0, \  |u_z| \to 0  \quad \text{as} \   |z|  \to \infty.
\label{BC}
\ee

Central to our approach is the observation that 
the equation \eqref{A1}
can be written as a first-order system
\begin{align}
u_z -  & iA u+ v=0,   \nonumber \\ 
v_z+  &  i A v - 2u|u|^2+ u=0.  \nonumber  
\end{align} 
The polar decomposition 
\[
u= a e^{i \theta}, \quad
v= b e^{i \chi},
\]
where $a>0$, $0 \leq \theta < 2 \pi$ and $b^2>0$, $0 \leq \chi < \pi$,
takes this system to
\begin{subequations}
\begin{align}
a_z & = - b \cos \mu,            \label{A2} \\
b_z  &  = a({ 2}a^2-1) \cos \mu,  \label{A3}    \\
(\theta_z-A)a   & = - b \sin \mu,  \label{A4}         \\
(\chi_z+A) b  & = a(1-{ 2} a^2) \sin \mu,  \label{A5}
\end{align}
\end{subequations} 
where we have introduced the angle
\[
\mu(z)  = \chi(z) -\theta(z).
\]

An immediate consequence of equations \eqref{A2}-\eqref{A3} is a conservation law
\[
a^2 \left( 1-a^2 \right) = b^2+C,
\]
where 
 $C$ is a constant. 
Equation \eqref{A4}, along with the boundary conditions \eqref{BC} 
and the fact that $A$  remains bounded as $|z| \to \infty$, 
 gives $b \sin \mu \to 0$. On the other hand,  equation \eqref{A2} implies
$b \cos \mu \to 0$. Taken together, these two results lead us to  conclude that  $b \to 0$
as $|z| \to \infty$ and so $C=0$:
\be
a^2 \left( 1-a^2 \right) = b^2.
\label{A6}
\ee

With the relation \eqref{A6} in place, equation \eqref{A2} 
can be integrated to give
\be
a=  {\rm sech} (\Phi- \Phi_0), \quad 
b=   {\rm sech}   (\Phi- \Phi_0) {\rm tanh} ( \Phi- \Phi_0),    
\label{A7}
\ee
where
\[
\Phi(z)= \int_0^z \cos \mu (s) ds, 
\]
and $\Phi_0$ is a constant of integration. 
The remaining two equations, \eqref{A4} and \eqref{A5}, can be solved for
$\theta$ and $A$:
\begin{align}
\theta= -\frac{\mu}{2} - \frac12 \int \frac{a^3}{b}  \sin \mu dz,  \label {A9} \\
A=  -\frac{\mu_z}{2} +  \frac{a}{2b} \left(2- 3a^2 \right) \sin \mu.  \label{A10}
\end{align}

The seed function $\cos \mu(z)$ can be chosen arbitrarily. Once $\mu(z)$  has been
chosen, the first equation in 
\eqref{A7} gives $a(z)$
while equation \eqref{A9} together with the second equation in \eqref{A7}
produce
 $\theta(z)$. 
 The corresponding potential base function $A(z)$
is given by  \eqref{A10}. 


In this contribution we confine ourselves to the seed functions whose integrals $\Phi(z)$ 
 are bounded (from above or from below)
over the whole line. 
Assuming, for definiteness, that $\Phi(z)$ is bounded from below and
choosing the constant $\Phi_0$ to satisfy
\[
\Phi_0< \inf_{-\infty< z< \infty} \Phi(z),
\]
we will 
ensure that $\Phi(z)-\Phi_0>0$ and
 the function $b(z)$ in \eqref{A7} is bounded away from zero.
  Then the quotient
$a(z)/b(z)$  in \eqref{A9} and \eqref{A10} will be nonsingular:
\[
\left| \frac{a}{b} \right|  = {\rm cotanh}  \left| \Phi(z)-\Phi_0 \right| < \infty.
\]
A simple class of suitable $\Phi(z)$ consists of 
even functions  bounded by their value at the origin.

Finally,  equation \eqref{A7} implies that the solution will only be localised (that is, satisfy the boundary conditions \eqref{BC})
if the integral $\int_{-\infty}^{\infty}  \cos \mu \, ds$ diverges.
This means that  $\cos \mu(z)$ should either remain nonzero as $|z| \to \infty$
(for example, tend to a nonzero constant), or
decay  to zero --- but no faster than $z^{-1}$.

\section{Pulse-like solitons: two simple examples}
\label{2ex} 

As our first example we take the seed function of the form
\be
\cos \mu(z) = \frac{\sinh z}{\sqrt{ \sinh^2 z+ \cos^2 \alpha}},
\label{seat}
\ee
where 
$0 \leq  \alpha< \pi/2$ is a parameter. The corresponding integral
\be
\Phi(z) = {\rm Arctanh} \sqrt{1-\sin^2 \alpha  \, \sech^2z}
\label{B4}
\ee
is even and monotonically growing from ${\rm Arctanh} (\cos \alpha)$ to infinity 
as $z$ varies from 0 to $\infty$.  We let $\Phi_0=0$, for simplicity.

Equations \eqref{A7}, 
 \eqref{A9},  and \eqref{A10} give
 the potential base function
 \be
 A= \frac32  \cos{\alpha} \, \sech \,  z,  \label{B3}
\ee 
as well as the absolute value  and phase of the soliton:
\be
a=  \sin \alpha  \, \sech \, z, \quad
\theta= \frac12 \cos \alpha  \arctan \left( \sinh  \, z \right). \label{B2}
\ee
The Wadati potential $-A^2+iA_z$ with $A$ as in \eqref{B3} belongs to the class 
 of  \PT-symmetric potentials considered by Musslimani et al \cite{Ziad},
and equations \eqref{B2} constitute their soliton solution. 

Our second example is equally simple --- yet new. 
This time the seed function is 
\[
\cos \mu(z) = \frac{1}{\sqrt{1+z^2}},
\]
with the integral
\[
\Phi(z) = {\rm Arctanh} \frac{z}{\sqrt{1+z^2}}.
\]
The function  $\Phi(z)$ grows without bound as $z$ changes from zero to infinity. 
Making use of equations  \eqref{A7}, 
 \eqref{A9},  and \eqref{A10} we arrive at the base 
 \[
 A(z)= 1- \frac{2}{1+z^2}
 \]
and the corresponding  \PT-symmetric complex Wadati potential:
  \be
-A^2+iA_z= -1 +  \frac{4z(z+i)}{(z^2+1)^2}.
 \label{alg1}
 \ee 
 The localised nonlinear mode, or the soliton, supported by this potential is also given by a rational function:
 \be
 u(z)= \frac{1-iz}{1+z^2}.
 \label{alg2}
 \ee
 (We remind the reader that $z$ is a real coordinate in \eqref{alg1} and \eqref{alg2}.)
 
\begin{figure}[t]
\center
\includegraphics[width=8cm]{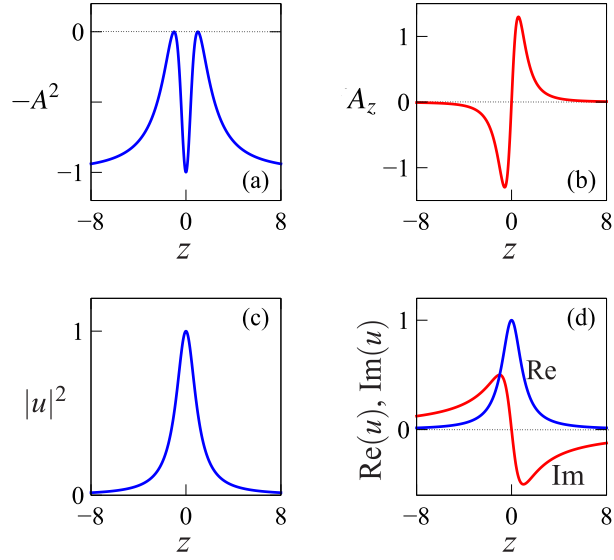}
\caption{
Top row: the real (a) and imaginary (b) part of the rational-function $\PT$-symmetric potential (\ref{alg1}).
Bottom row:  the modulus squared of the corresponding localised nonlinear mode (c) 
and its real and imaginary part (d). 
 }
\label{Fig1}       
\end{figure}

 The  potential (\ref{alg1}) and the soliton (\ref{alg2}) are depicted in Fig.~\ref{Fig1}.

\section{Repulsive nonlinearity and nonvanishing backgrounds}
\label{Dark}

Turning to the Gross-Pitaevskii equation with repulsion
(equation \eqref{A0} with $g<0$), we focus on solitons in the constant-density condensate, that is, localised solutions 
satisfying the nonvanishing boundary conditions at infinity:
\be
|\psi(x)|^2 \to \rho_0, \, |\psi_x(x)| \to 0
\quad \text{as} \ |x| \to \infty.
\label{BC0}
\ee
Assuming that the potential has been gauged so that $W(x) \to 0$ as $|x| \to \infty$, 
the conditions \eqref{BC0} require $\kappa^2=g\rho_0 <0$.
Scaling the dependent and independent variables as in
\[
A(z)= \lambda W(x), \quad 
u(z)= \sqrt{-\frac{g}{2} } \lambda \psi(x),
\quad
z= \frac{x}{\lambda},
\]
with $\lambda=\sqrt{-2/\kappa^2}>0$,
the equation \eqref{A0} becomes
\be
u_{zz} + (A^2- iA_z) u  -2 u|u|^2 =  -2u,
\label{C1}
\ee
while the nonvanishing boundary conditions are reduced to
\be
|u|^2 \to 1, \ |u_z|^2 \to 0 \quad \text{as} \ |z| \to \infty.
\label{BC2}
\ee

The equation \eqref{C1} can be written as a first-order system
\begin{align}
u_z -  & iA u  + v  =0,     \label{D3} \\
v_z+  &  i A v +2 u|u|^2- 2 u=0.    \label{D4}
\end{align} 
In the same way as the system \eqref{A2}-\eqref{A3} gave rise to the conservation law \eqref{A6}, 
the system \eqref{D3}-\eqref{D4} implies
\be
b= \pm (1-a^2),   \label{C7}
\ee
where we have introduced the polar decomposition 
\[
u= a e^{i \theta}, \quad
v= b e^{i \chi},
\]
and used
the boundary conditions \eqref{BC2}
together with the fact that $A \to 0$ at infinity.
 In \eqref{C7},  the top sign  corresponds to solutions with $a \leq 1$ while
the bottom sign  pertains to those with $a \geq 1$. 

Letting  $\mu(z)  = \chi(z) -\theta(z)$ and
making use of the conservation law \eqref{C7} we obtain the modulus and phase of the solution in the sector $a \leq 1$:
\be
a= -  \tanh  ( \Phi-\Phi_0),  \quad  \theta= -\frac{\mu}{2}  -  \int \frac{(a^2+1)}{2a} \sin \mu \, dz.
\label{C8} 
\ee
Here
\[
\Phi(z) = \int_0^z \cos \mu(s) ds,
\]
as before. 
The Wadati potential bearing the solution \eqref{C8} is based on the function
\be
 A= -\frac{\mu_z}{2}    + \frac{(1-3a^2) }{2a} \sin \mu.  \label{C13}
 \ee
 
In the sector $a \geq 1$, 
 the potential base and solution  are given by
\begin{align*} 
A= -\frac{\mu_z }{2}   + \frac{(3a^2-1)}{2a} \sin \mu;  \\
a=   \cotanh  ( \Phi-\Phi_0), \quad
\theta= -\frac{\mu}{2}  +   \int \frac{(a^2+1)}{2a} \sin \mu \, dz.
\end{align*}

\section{Lumps and bubbles in a homogeneous condensate}
\label{kinks} 

Consider, first, the case $a \leq 1$ and let
\be
\cos \mu(z) = - \frac{\sinh z}{\sqrt{\sinh^2z + \cos^2 \alpha}},
\label{cosbub}
\ee
where $0 \leq \alpha < \pi/2$ is a parameter. (This is a  negative of the 
seed function \eqref{seat} employed in the attractive case.)
Using the integral
\[
\Phi(z)  = -{\rm Arctanh} \,\sqrt{1-\sin^2 \alpha  \, \sech^2z}
\]
  and letting $\Phi_0=0$,
equation \eqref{C13} provides the potential base:
\be
A=  \frac{3 \cos \alpha}{2}  \, \sech \, z.
\label{bub2}
\ee
Equations \eqref{C8} give the
corresponding solution:
\be
u=  (\cos \alpha \, \sech \, z+ i \tanh z)(\sech \, z+ i \tanh z)^\sigma,
\quad 
\sigma= \frac{\cos \alpha}{2}.
\label{bub1}
\ee

The quantity $|u|^2$ has a dip at the origin:
\[
|u|^2= 1- \sin^2{\alpha} \, \sech^2 z.
\]
Therefore, equation \eqref{bub1} describes  a bubble ---  a localised rarefaction in a homogeneous background density.\footnote{
In nonlinear dynamics, the   {\it bubble\/} refers to  a particular class of nontopological solitons with 
nontrivial boundary conditions \cite{bubble1,bubble2}. In contrast to the  strict mathematical terminology, we use this word
in a broad physical sense here --- as a synonym of a hole in the constant-density condensate. The optical equivalent of 
the condensate bubble is  {\it dark soliton}.}

In the sector $a \geq 1$, choosing
\[
\cos \mu(z)= \frac{\sinh z}{\sqrt{\sinh^2z + \cosh^2 \beta}}, 
\quad
\Phi(z)= {\rm Arc coth} \sqrt{1+ \sinh^2 \beta \, \sech^2 z},
\]
with $\beta$ a real parameter, $0 \leq \beta < \infty$, gives rise to the potential base 
\be
A= \frac{3 \cosh \beta}{2}  \sech \, z
\label{tw1}
\ee
and the solution 
\[
u= (\cosh \beta \, \sech \,  z+ i \tanh z)(\sech  \, z + i \tanh z)^\sigma, 
\quad \sigma= \frac{\cosh \beta}{2}.
\]
This time the quantity $|u|^2$ has a  maximum at the origin,
\[
|u|^2= 1+ \sinh^2 \beta \, \sech^2 z,
\]
and so the solution describes a lump --- a localised domain of compression in a condensate of uniform density.

Note that the potential base \eqref{bub2} can be  formally obtained from \eqref{tw1}  by letting $\beta= i \alpha$.
Therefore the bubble- and lump-like solitons
form a seamless one-parameter family.

\section{Solitons in a stationary flow}
\label{flow}

In  the boson-condensate  interpretation of solutions to the equation \eqref{A1}, the function $J(z)=a^2 \theta_z $ represents
 the superfluid current. Physically, of interest are  stationary flows, that is, solutions with $J(z)$ approaching  nonzero values  as $z \to\pm \infty$. 
 In this section we construct Wadati potentials supporting  the stationary flow of condensate.

The lump and the bubble solitons from the previous section
are characterised by the zero current at  infinity. To construct exact solutions with a nonzero stationary current,  we modify the seed function \eqref{cosbub}
by introducing an additional parameter:
\[
\cos \mu  = -           \sin \varphi \,    \frac   {\sinh  y }{\sqrt {\sinh^2 y+ \cos^2 \alpha
			 }},
			\quad
			y= \sin (\varphi) z.
		\]
Here $0 < \varphi \leq \pi/2$. 
The integral of the seed is
\[
\Phi(z)   = -{\rm Arctanh} \,\sqrt{1-\sin^2 \alpha \,  \sech^2 y}.	
\]

The corresponding \PT-symmetric Wadati potential is generated by the  base function
\begin{equation}
A(z) = \frac{\, 3 (\sin^2 \alpha-\sin^2 \varphi)
	-2\cos^2 \varphi   \cosh^2  {y}}
{2  \cosh  {y}
	   \sqrt {\cos^2 \varphi    \sinh^2  y
				+ \cos^2 \alpha
				 }}.
				 \label{Af}
\end{equation}
The absolute value of the corresponding
 solution has a simple form,
 \begin{subequations}
\be
a(z) = \sqrt{1-\sin^2 \alpha \, \sech^2y},
\label{abs}
\ee
and the phase gradient is given by
\begin{equation}
\theta_z = \frac{
  (\sin^2 \varphi- \sin^2 \alpha) (\sin^2 \alpha - 3 \cosh^2 y)-2\cos^2 \varphi \cosh^4 y
  } 
{2  \cosh{y}
	 \left(  \cosh^2
 {y}   -\sin^{2} \alpha \right) {{ \sqrt{\cos^2 \varphi    \sinh^2  y
				+ \cos^2 \alpha
				 				}}} }.
\label{tf}
\end{equation}
\label{flowing} 
\end{subequations}

When 
$ \varphi \neq  \pi/2$,  the solution represents a stationary flow:
\[
\left. J \right|_{z \to \pm \infty}
 = -\cos \varphi \neq  0.
\]
Varying $\varphi$ we can generate potential-solution pairs with negative
 currents ranging  from $-1$ to 0.
(Inserting a minus in front of the right-hand side in \eqref{Af}
and \eqref{tf} produces  pairs with  positive currents.)

\begin{figure}[t]
\center
\includegraphics[width=9cm]{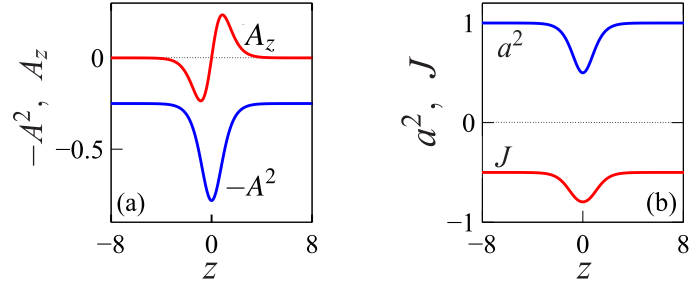}
%
%
\caption{(a) The real (blue) and imaginary (red) part of the 
\PT-symmetric Wadati potential   generated by the  base function (\ref{Af}).
 (b) The corresponding nonlinear mode \eqref{flowing}.
 Shown is the condensate density $a^2$ and the superfluid current $J=a^2\theta_z$. In these plots,  $\alpha=\pi/4$ and $\varphi=\pi/3$.
}
\label{Fig2}       
\end{figure}

An example of the potential generated by the base function (\ref{Af}) and the corresponding nonlinear mode are shown in Fig.~\ref{Fig2}.

\section{Concluding remarks}
\label{conclusions}

Due to their unique properties, Gross-Pitaevskii equations with the Wadati potentials are of particular interest 
in \PT-symmetric theories.  Accordingly, it would be desirable to have a 
sufficiently diverse and ample collection of {\it exactly-solvable\/} Wadati potentials --- these would serve as testing grounds 
for realistic physical models and starting points for perturbation expansions. The purpose of this contribution was to show how one can generate broad classes of 
\PT-symmetric Wadati potentials along with exact localised solutions of the associated Gross-Pitaevskii equations.

The crux of our method lies in the ability to write the nonlinear second-order equation with a Wadati potential, as 
a symmetric system of two first-order equations. The potential function of this first-order system is nothing but the 
base function of the Wadati potential of the original second-order equation. 

A practically-minded reader may naturally wonder what is the advantage of our approach over a simple  reverse engineering,
where the potential $V(x)$ is reconstructed from a postulated localised solution of equation \eqref{A0}:
\be
V(x)= \frac{\psi_{xx}}{\psi} + g |\psi|^2 -\kappa^2.
\label{ri}
\ee
The answer is that the back-engineered potential \eqref{ri} will generally \emph{not\/} be of the Wadati variety. 

In contrast, our method constructs the {\it base function\/} first. Only after the base $W(x)$ has been 
constructed does one proceed to form the potential $V=-W^2+iW_x$. Thus the resultant potential is Wadati by construction.

We have exemplified this procedure by constructing several  exactly
solvable \PT-symmetric Wadati potentials for the attractive and repulsive Gross-Pitaevskii. In the case of the attractive
(``focussing") cubic nonlinearity, equation \eqref{B3} reproduces the base function known in literature
while the rational  potential \eqref{alg1} is new.
In the repulsive (``defocussing") situation, the bases \eqref{bub2} and \eqref{tw1} constitute a continuous family of  Wadati potentials
supporting solitons over a nonvanishing background (lumps and bubbles). To the best of our knowledge, these potential-solution pairs are also new. 
Finally, we have constructed an exactly-solvable \PT-symmetric potential supporting bubble-like
solitons in a stationary flow of the superfluid. The corresponding potential base function is in \eqref{Af}.

\begin{acknowledgement}

 This work was supported by the NRF of South Africa (grants UID 85751, 86991, and 87814) and the FCT (Portugal) through the grants UID/FIS/00618/2013 and PTDC/FIS-OPT/1918/2012.
 One of the authors (IVB) also thanks the Israel Institute for Advanced Studies for partial financial support.
 \end{acknowledgement}
%

%
%
%

\end{document}